\catcode`\@=11

\font\tenmsa=msam10
\font\sevenmsa=msam7
\font\fivemsa=msam5
\font\tenmsb=msbm10
\font\sevenmsb=msbm7
\font\fivemsb=msbm5
\newfam\msafam
\newfam\msbfam
\textfont\msafam=\tenmsa  \scriptfont\msafam=\sevenmsa
  \scriptscriptfont\msafam=\fivemsa
\textfont\msbfam=\tenmsb  \scriptfont\msbfam=\sevenmsb
  \scriptscriptfont\msbfam=\fivemsb

\def\hexnumber@#1{\ifnum#1<10 \number#1\else
 \ifnum#1=10 A\else\ifnum#1=11 B\else\ifnum#1=12 C\else
 \ifnum#1=13 D\else\ifnum#1=14 E\else\ifnum#1=15 F\fi\fi\fi\fi\fi\fi\fi}

\def\msa@{\hexnumber@\msafam}
\def\msb@{\hexnumber@\msbfam}
\mathchardef\boxdot="2\msa@00
\mathchardef\boxplus="2\msa@01
\mathchardef\boxtimes="2\msa@02
\mathchardef\square="0\msa@03
\mathchardef\blacksquare="0\msa@04
\mathchardef\centerdot="2\msa@05
\mathchardef\lozenge="0\msa@06
\mathchardef\blacklozenge="0\msa@07
\mathchardef\circlearrowright="3\msa@08
\mathchardef\circlearrowleft="3\msa@09
\mathchardef\rightleftharpoons="3\msa@0A
\mathchardef\leftrightharpoons="3\msa@0B
\mathchardef\boxminus="2\msa@0C
\mathchardef\Vdash="3\msa@0D
\mathchardef\Vvdash="3\msa@0E
\mathchardef\vDash="3\msa@0F
\mathchardef\twoheadrightarrow="3\msa@10
\mathchardef\twoheadleftarrow="3\msa@11
\mathchardef\leftleftarrows="3\msa@12
\mathchardef\rightrightarrows="3\msa@13
\mathchardef\upuparrows="3\msa@14
\mathchardef\downdownarrows="3\msa@15
\mathchardef\upharpoonright="3\msa@16

\mathchardef\downharpoonright="3\msa@17
\mathchardef\upharpoonleft="3\msa@18
\mathchardef\downharpoonleft="3\msa@19
\mathchardef\rightarrowtail="3\msa@1A
\mathchardef\leftarrowtail="3\msa@1B
\mathchardef\leftrightarrows="3\msa@1C
\mathchardef\rightleftarrows="3\msa@1D
\mathchardef\Lsh="3\msa@1E
\mathchardef\Rsh="3\msa@1F
\mathchardef\rightsquigarrow="3\msa@20
\mathchardef\leftrightsquigarrow="3\msa@21
\mathchardef\looparrowleft="3\msa@22
\mathchardef\looparrowright="3\msa@23
\mathchardef\circeq="3\msa@24
\mathchardef\succsim="3\msa@25
\mathchardef\gtrsim="3\msa@26
\mathchardef\gtrapprox="3\msa@27
\mathchardef\multimap="3\msa@28
\mathchardef\therefore="3\msa@29
\mathchardef\because="3\msa@2A
\mathchardef\doteqdot="3\msa@2B

\mathchardef\triangleq="3\msa@2C
\mathchardef\precsim="3\msa@2D
\mathchardef\lesssim="3\msa@2E
\mathchardef\lessapprox="3\msa@2F
\mathchardef\eqslantless="3\msa@30
\mathchardef\eqslantgtr="3\msa@31
\mathchardef\curlyeqprec="3\msa@32
\mathchardef\curlyeqsucc="3\msa@33
\mathchardef\preccurlyeq="3\msa@34
\mathchardef\leqq="3\msa@35
\mathchardef\leqslant="3\msa@36
\mathchardef\lessgtr="3\msa@37
\mathchardef\backprime="0\msa@38
\mathchardef\risingdotseq="3\msa@3A
\mathchardef\fallingdotseq="3\msa@3B
\mathchardef\succcurlyeq="3\msa@3C
\mathchardef\geqq="3\msa@3D
\mathchardef\geqslant="3\msa@3E
\mathchardef\gtrless="3\msa@3F
\mathchardef\sqsubset="3\msa@40
\mathchardef\sqsupset="3\msa@41
\mathchardef\trianglerighteq="3\msa@44
\mathchardef\trianglelefteq="3\msa@45
\mathchardef\bigstar="0\msa@46
\mathchardef\between="3\msa@47
\mathchardef\blacktriangledown="0\msa@48
\mathchardef\blacktriangleright="3\msa@49
\mathchardef\blacktriangleleft="3\msa@4A
\mathchardef\blacktriangle="0\msa@4E
\mathchardef\triangledown="0\msa@4F
\mathchardef\eqcirc="3\msa@50
\mathchardef\lesseqgtr="3\msa@51
\mathchardef\gtreqless="3\msa@52
\mathchardef\lesseqqgtr="3\msa@53
\mathchardef\gtreqqless="3\msa@54
\mathchardef\Rrightarrow="3\msa@56
\mathchardef\Lleftarrow="3\msa@57
\mathchardef\veebar="2\msa@59
\mathchardef\barwedge="2\msa@5A
\mathchardef\doublebarwedge="2\msa@5B
\mathchardef\angle="0\msa@5C
\mathchardef\measuredangle="0\msa@5D
\mathchardef\sphericalangle="0\msa@5E
\mathchardef\varpropto="3\msa@5F
\mathchardef\smallsmile="3\msa@60
\mathchardef\smallfrown="3\msa@61
\mathchardef\Subset="3\msa@62
\mathchardef\Supset="3\msa@63
\mathchardef\Cup="2\msa@64

\mathchardef\Cap="2\msa@65

\mathchardef\curlywedge="2\msa@66
\mathchardef\curlyvee="2\msa@67
\mathchardef\leftthreetimes="2\msa@68
\mathchardef\rightthreetimes="2\msa@69
\mathchardef\subseteqq="3\msa@6A
\mathchardef\supseteqq="3\msa@6B
\mathchardef\bumpeq="3\msa@6C
\mathchardef\Bumpeq="3\msa@6D
\mathchardef\lll="3\msa@6E

\mathchardef\ggg="3\msa@6F

\mathchardef\circledS="0\msa@73
\mathchardef\pitchfork="3\msa@74
\mathchardef\dotplus="2\msa@75
\mathchardef\backsim="3\msa@76
\mathchardef\backsimeq="3\msa@77
\mathchardef\complement="0\msa@7B
\mathchardef\intercal="2\msa@7C
\mathchardef\circledcirc="2\msa@7D
\mathchardef\circledast="2\msa@7E
\mathchardef\circleddash="2\msa@7F
\def\ulcorner{\delimiter"4\msa@70\msa@70 }
\def\urcorner{\delimiter"5\msa@71\msa@71 }
\def\llcorner{\delimiter"4\msa@78\msa@78 }
\def\lrcorner{\delimiter"5\msa@79\msa@79 }
\def\yen{\mathhexbox\msa@55 }
\def\checkmark{\mathhexbox\msa@58 }
\def\circledR{\mathhexbox\msa@72 }
\def\maltese{\mathhexbox\msa@7A }
\mathchardef\lvertneqq="3\msb@00
\mathchardef\gvertneqq="3\msb@01
\mathchardef\nleq="3\msb@02
\mathchardef\ngeq="3\msb@03
\mathchardef\nless="3\msb@04
\mathchardef\ngtr="3\msb@05
\mathchardef\nprec="3\msb@06
\mathchardef\nsucc="3\msb@07
\mathchardef\lneqq="3\msb@08
\mathchardef\gneqq="3\msb@09
\mathchardef\nleqslant="3\msb@0A
\mathchardef\ngeqslant="3\msb@0B
\mathchardef\lneq="3\msb@0C
\mathchardef\gneq="3\msb@0D
\mathchardef\npreceq="3\msb@0E
\mathchardef\nsucceq="3\msb@0F
\mathchardef\precnsim="3\msb@10
\mathchardef\succnsim="3\msb@11
\mathchardef\lnsim="3\msb@12
\mathchardef\gnsim="3\msb@13
\mathchardef\nleqq="3\msb@14
\mathchardef\ngeqq="3\msb@15
\mathchardef\precneqq="3\msb@16
\mathchardef\succneqq="3\msb@17
\mathchardef\precnapprox="3\msb@18
\mathchardef\succnapprox="3\msb@19
\mathchardef\lnapprox="3\msb@1A
\mathchardef\gnapprox="3\msb@1B
\mathchardef\nsim="3\msb@1C
\mathchardef\napprox="3\msb@1D
\mathchardef\nsubseteqq="3\msb@22
\mathchardef\nsupseteqq="3\msb@23
\mathchardef\subsetneqq="3\msb@24
\mathchardef\supsetneqq="3\msb@25
\mathchardef\subsetneq="3\msb@28
\mathchardef\supsetneq="3\msb@29
\mathchardef\nsubseteq="3\msb@2A
\mathchardef\nsupseteq="3\msb@2B
\mathchardef\nparallel="3\msb@2C
\mathchardef\nmid="3\msb@2D
\mathchardef\nshortmid="3\msb@2E
\mathchardef\nshortparallel="3\msb@2F
\mathchardef\nvdash="3\msb@30
\mathchardef\nVdash="3\msb@31
\mathchardef\nvDash="3\msb@32
\mathchardef\nVDash="3\msb@33
\mathchardef\ntrianglerighteq="3\msb@34
\mathchardef\ntrianglelefteq="3\msb@35
\mathchardef\ntriangleleft="3\msb@36
\mathchardef\ntriangleright="3\msb@37
\mathchardef\nleftarrow="3\msb@38
\mathchardef\nrightarrow="3\msb@39
\mathchardef\nLeftarrow="3\msb@3A
\mathchardef\nRightarrow="3\msb@3B
\mathchardef\nLeftrightarrow="3\msb@3C
\mathchardef\nleftrightarrow="3\msb@3D
\mathchardef\divideontimes="2\msb@3E
\mathchardef\varnothing="0\msb@3F
\mathchardef\nexists="0\msb@40
\mathchardef\mho="0\msb@66
\mathchardef\thorn="0\msb@67
\mathchardef\beth="0\msb@69
\mathchardef\gimel="0\msb@6A
\mathchardef\daleth="0\msb@6B
\mathchardef\lessdot="3\msb@6C
\mathchardef\gtrdot="3\msb@6D
\mathchardef\ltimes="2\msb@6E
\mathchardef\rtimes="2\msb@6F
\mathchardef\shortmid="3\msb@70
\mathchardef\shortparallel="3\msb@71
\mathchardef\smallsetminus="2\msb@72
\mathchardef\thicksim="3\msb@73
\mathchardef\thickapprox="3\msb@74
\mathchardef\approxeq="3\msb@75
\mathchardef\succapprox="3\msb@76
\mathchardef\precapprox="3\msb@77
\mathchardef\curvearrowleft="3\msb@78
\mathchardef\curvearrowright="3\msb@79
\mathchardef\digamma="0\msb@7A
\mathchardef\varkappa="0\msb@7B
\mathchardef\hslash="0\msb@7D
\mathchardef\hbar="0\msb@7E
\mathchardef\backepsilon="3\msb@7F
\def\Bbb{\ifmmode\let\next\Bbb@\else
 \def\next{\errmessage{Use \string\Bbb\space only in math mode}}\fi\next}
\def\Bbb@#1{{\Bbb@@{#1}}}
\def\Bbb@@#1{\fam\msbfam#1}

\catcode`\@=\active

\def\R{{\Bbb R}}
\def\C{{\Bbb C}}

\def\<{\langle}
\def\>{\rangle}
\def\del{{\partial}}

\def\eps{{\epsilon}}

\def\rcross{{\triangleright\!\!\!<}}

\def\cobicross{{\triangleright\!\!\!\blacktriangleleft}}
\def\bicross{{\blacktriangleright\!\!\!\triangleleft}}
\def\dcross{{\bowtie}}

\def\cosub{{\Delta\kern -.65em \raisebox{.02em}{-}\kern .35em}}

\def\tens{\mathop{\otimes}}
\def\la{{\triangleright}}\def\ra{{\triangleleft}}

\def\id{{\rm id}}

\def\o{{}_{\scriptscriptstyle(1)}}
\def\t{{}_{\scriptscriptstyle(2)}}
\def\th{{}_{\scriptscriptstyle(3)}}

\def\new#1{\goodbreak\goodbreak\bigskip\noindent{\bf #1}}
\def\note#1{}

\def\equad{\kern -1.7em}

\def\eqn#1#2{\begin{equation}#2\label{#1}\end{equation}}

\documentstyle[11pt]{article}
\def\kp{{P_\kappa}}
\def\newN{{\cal N}} 
\def\newP{{\cal P}} 

\begin{document}\baselineskip 21pt

{\ }\hskip 4.2in DAMTP/94-24

{\ }\hskip 4.2in UGVA-DPT 1994/03-844
\vspace{.5in}

\begin{center} {\LARGE BICROSSPRODUCT STRUCTURE OF $\kappa$-POINCARE GROUP AND
NON-COMMUTATIVE GEOMETRY}
\\ \baselineskip 13pt{\ }
{\ }\\ S. Majid
\footnote{SERC Fellow and Fellow of Pembroke College,
Cambridge}\\ {\ }\\
Department of Applied Mathematics \& Theoretical Physics\\ University of
Cambridge, Cambridge CB3 9EW, U.K.\\
{\ }\\H. Ruegg
\footnote{Partially supported by the Swiss National Science Foundation}
\\ {\ }\\
D\'epartement de Physique th\'eorique \\ Universit\'e de Gen\`eve \\
CH-1211 Gen\`eve 4, Switzerland \\
\end{center}

\begin{center}
March 1994\end{center}
\vspace{10pt}
\begin{quote}\baselineskip 13pt
\noindent{\bf ABSTRACT} We show that the $\kappa$-deformed Poincar\'e quantum
algebra proposed for elementary particle physics has the structure of a Hopf
algebra bicrossproduct $U(so(1,3))\cobicross T$. The algebra is a semidirect
product of the classical Lorentz group $so(1,3)$ acting in a deformed way on
the momentum sector $T$. The novel feature is that the coalgebra is also
semidirect, with a backreaction of the momentum sector on the Lorentz
rotations. Using this, we show that the $\kappa$-Poincare acts covariantly on a
$\kappa$-Minkowski space, which we introduce. It turns out necessarily to be
deformed and non-commutative. We also connect this algebra with a previous
approach to Planck scale physics.
\end{quote}
\baselineskip 21pt

\new{1} This is a note on the $\kappa$-Poincare algebra as introduced in
\cite{LNRT}\cite{LNR} and studied extensively with a view to applications
in
elementary particle physics\cite{N}\cite{LRR}\cite{B}\cite{D}\cite{R}. The idea
behind this particular
deformation, which is
obtained by contraction\cite{C}, is that it is one of the weakest possible
deformations
of the usual
Poincare group as a Hopf algebra. Hence it provides an ideal testing-ground for
possible applications in particle physics. The momenta remain commutative
\eqn{PP}{ [P_\mu,P_\nu] = 0}
and the rotation part of the Lorentz sector is also not deformed. Because of
the mildness of the deformation, many particle constructions and predictions
can be obtained easily.

Here we want to argue that in spite of this success, any application of the
$\kappa$-Poincare group to physics leads necessarily into non-commutative
geometry. This is because until now it has not been possible to define an
algebra of Minkowski space co-ordinates $\{x_\mu\}$ on which the
$\kappa$-Poincare acts as a Hopf algebra. Recall that when usual groups act on
algebras, one has
\eqn{group}{ g\la (ab)=(g\la a)(g\la b),\quad g\la 1=1 }
and the natural analogue of this for Hopf algebras is
\eqn{covact}{ h\la(ab)=(h\o\la a)(h\t\la b),\quad h\la 1=\eps(h)1}
where $\Delta h=h\o\tens h\t=\sum_i{h_{(1)i}\tens h_{(2)i}}$ is the coproduct.
Without such a covariant
action, one cannot
make any products of the space-time generators $x_\mu$ in a $\kappa$-Poincare
invariant way. This affects not only the many-particle theory but any
expressions involving, for example, $x^2$. It means that until now, the actual
coproduct structure has only been applied in connection with momentum space and
not spacetime itself . Since the
coproduct of the $\kappa$-Poincare is non-cocommutative, one cannot expect that
it acts on the usual commutative algebra of functions on Minkowski space: it
needs to be non-commutative or `quantum'.

Here we provide the correct notion of $\kappa$-Minkowski space and the action
of $\kappa$-Poincare on it. We also understand the structure of the
$\kappa$-Poincare as a
deformation of the usual semidirect product structure. This then makes
tractable the problem of representing it covariantly on the $\kappa$-Minkowski.

The abstract structure of the $\kappa$-Poincare turns out to be an example of a
class of non-commutative non-cocommutative Hopf algebras (quantum groups)
introduced some years ago by the first author in an algebraic approach to
Planck-scale physics\cite{Ma:pla}\cite{Ma:pri}\cite{Ma:phy}. The context here
was quite different, namely the Hopf algebra of observables of a
quantum system rather than as a symmetry object. Thus we find in fact that the
$\kappa$-Poincare algebra $\kp$ has {\em two} different physical
interpretations, one as a
quantum symmetry group and the other as a quantised phase space. Thus, we find
\eqn{bicross}{\kp =U(so(1,3))\cobicross T= U(so(1,3))\cobicross \C(X)}
where in the first picture $T$ is the $\kappa$-deformed enveloping algebra of
the momentum sector of the Poincare. In the second picture it is the algebra of
functions of a classical but curved momentum part $X$ of phase space. This
second point of view is recalled briefly in the last section of this note.

\new{2.}
The $\kappa$-Poincare algebra $\kp$, (antihermitian generators of translations
$P_{\mu}$, rotations $M_i$ and boosts $\bar N_i$; $\kappa$ real;
 $i,j,k=1,2,3; \mu,\nu=0,1,2,3$) is \cite{LNR} :

\eqn{LNR1}{ [P_{\mu},P_{\nu}]=0, \quad [M_i,M_j]=\eps_{ijk}M_k,  }
\eqn{LNR2}{ [M_i,P_j]=\eps_{ijk} P_k, \quad [M_i,P_0]=0, }
\eqn{LNR3}{ [M_i,\bar N_j]=\eps_{ijk}\bar N_k, }
\eqn{LNR4}{ [\bar N_i,P_0]=P_i, \quad [\bar N_i,P_j]=\delta_{ij}\kappa
\sinh{P_0\over{\kappa}}, }
\eqn{LNR5}{ [\bar N_i,\bar N_j]=-\eps_{ijk}(M_k \cosh{P_0\over{\kappa}} -
{1\over{4\kappa^2}}P_k\vec P\vec M). }

The coproducts are given by:
\eqn{LNR6}{ \Delta P_0=P_0\tens 1 +1\tens P_0, \quad  \Delta M_i=M_i\tens 1
+1\tens M_i, }
\eqn{LNR7}{ \Delta P_i=P_i\tens e^{{P_0\over 2\kappa}}+e^{-{P_0\over
2\kappa}}\tens P_i, }
\eqn{LNR8}{ \Delta \bar N_i=\bar N_i\tens e^{P_0\over 2\kappa}+e^{-{P_0\over
2\kappa}}\tens
\bar N_i+ {\eps_{ijk}\over 2\kappa}(P_j\tens M_ke^{P_0\over
2\kappa}+e^{-{P_0\over 2\kappa}}M_j\tens P_k). }

The starting point of our structure theorem is the observation that
$\kp$ contains $T=\{P_\mu\}$ as a sub-Hopf algebra and projects onto
$U(so(1,3))$ also as a Hopf algebra map:

\eqn{exta}{ T{\buildrel i\over\hookrightarrow }\kp {\buildrel \pi\over\to}
U(so(1,3)).}
The map $\pi$ consists of setting $P_{\mu}=0$ and mapping $M_i$ and $\bar N_i$
to
their
classical counterparts in the Lorentz group. It is easy to see that $i,\pi$ are
classical counterparts as
\eqn{pi}{ \pi(M_i)=M_i,\quad \pi(\bar N_i)=N_i.}

To this, we add now the maps
\eqn{extb}{ T{\buildrel p\over\leftarrow }\kp {\buildrel
j\over\hookleftarrow} U(so(1,3)),\quad \pi\circ j=\id,\quad p\circ i=\id}
where $j$ is an algebra homomorphism and $p$ is a linear map which is a
coalgebra homomorphism
\eqn{extc}{ (p\tens p)\circ\Delta =\Delta\circ p,\quad \eps\circ p=\eps.}
Moreover,
\eqn{extd}{ (id\tens j)\circ\Delta=(\pi\tens\id)\circ\Delta \circ j}
which says that $j$ intertwines the coaction of $U(so(1,3))$ on itself by
$\Delta$ and its coaction on $\kp $ by $(\pi\tens\id)\circ\Delta$. Likewise
\eqn{exte}{ p(a)t=p(ai(t)),\quad a\in \kp ,\quad t\in T}
which says that $p$ intertwines the action of $T$ on itself by
right-multiplication, with its action on $\kp $ by $i$ and multiplication
in $\kp $.

Indeed, we define
\eqn{j}{ j(N_i)\equiv \newN_i=\bar N_i
e^{-{P_0\over2\kappa}}-{\eps_{ijk}\over
2\kappa}M_j P_k e^{-{
P_0\over 2\kappa}},\quad j(M_i)=M_i}
which one can show to be an algebra homomorphism. The new generators $\newN_i$
have
coproducts
\eqn{deltanewN}{ \Delta \newN_i=\newN_i\tens 1+e^{-{P_0\over\kappa}}\tens
\newN_i+{\eps_{ijk}\over
\kappa}P_je^{-{P_0\over 2\kappa}}\tens M_k}
after which (\ref{extd}) is clear. We also define $p$ as the map that sets
$M_i=\bar N_i=0$
and the properties (\ref{extc}), (\ref{exte}) are then clear.

Now, the data (\ref{exta})--(\ref{exte}) say precisely that $\kp $ is a
Hopf algebra extension of $U(so(1,3))$ by $T$. The general theory of Hopf
algebra extensions has been introduced in
\cite{Ma:phy} \cite{Ma:mat} \cite{Sin:ext} (the
latter two covered the general case) and one knows that such extensions are
semidirect products. There is also the possibility of cocycles but these vanish
when $j$ is an algebra homomorphism and $p$ a coalgebra one, as in our case.
We deduce from this theory that (a) the classical Lorentz algebra acts on $T$
from the right by
\eqn{extf}{ t\ra h= j(Sh\o)tj(h\t),\quad \forall t\in T,\quad h\in U(so(1,3))}
and (b) $T$ coacts back on $U(so(1,3))$ from the left by
\eqn{extg}{ \beta(\pi(a))=p(a\o)Sp(a\th)\tens \pi(a\t),\quad \forall \pi(a)\in
U(so(1,3)).}
In both formulae $S$ denotes the appropriate antipode while $\Delta^2a=a\o\tens
a\t\tens a\th$ in the second formula. In both cases, the formulae are not
obviously well-defined, but $t\ra h$ as stated necessarily lies in (the image
under $i$ of) $T$, while $\beta$ does not depend on  $a\in \kp $ but only
its image $\pi(a)$.

In our case we have
\eqn{TLa}{ P_0\ra M_i=0,\quad P_i\ra M_j=\eps_{ijk} P_k,\quad P_0\ra
N_i=-P_i e^{-{P_0\over2\kappa}}\equiv -\newP_i,}

the generators $\newP_i=P_i e^{-{P_0\over 2\kappa}}$ are
quite natural here, and in terms of these the action becomes

\eqn{TLb}{  \quad \newP_i\ra M_j=\eps_{ijk}\newP_k,\quad \newP_i\ra
N_j=-\delta_{ij}({\kappa \over
2}(1-e^{-{2P_0\over\kappa}})+{1\over 2\kappa}{\vec \newP}^2)+{1\over
\kappa}\newP_i\newP_j }
as computed for other reasons in \cite{RT}. Our present point of view
is not that this is the quantum adjoint action in $\kp $ but simply that
the classical $U(so(1,3))$ acts on the Hopf algebra $T$ in this way. Meanwhile,
the coaction comes out as
\eqn{betaL}{ \beta(M_i)=1\tens M_i,\quad \beta(N_i)=e^{-{P_0\over\kappa}}\tens
N_i+{\eps_{ijk}\over\kappa}\newP_j\tens M_k}
on the generators. Here $\beta$ is not an algebra homomorphism but its values
on products of generators can be computed too from (\ref{extg}).

Finally, the general extension theory says that our $\kp $ is built up in
its structure from this data $(T,U(so(1,3)),\ra,\beta)$. Namely, its algebra is
a semidirect product defined abstractly by $i(T)$ and $j(U(so(1,3))$ as
subalgebras and cross relations
\eqn{exth}{ i(t)j(h)=j(h\o)i(t\ra h\t),\quad \forall h\in U(so(1,3)),\quad t\in
T.}
Its coalgebra is defined in a dual way as
\eqn{exti}{ \Delta i(t)=i(t\o)\tens i(t\t),\quad
 \Delta j(h)= j(h\o)(i\tens j)\circ\beta(h\t).}

In our case the cross relations become

\eqn{crossTL}{ [P_0,M_i]=P_0\ra M_i,\quad [\newP_i,M_j]=\newP_i\ra M_j,\quad
[P_0,\newN_i]=P_0\ra
N_i,\quad [\newP_i,\newN_j]=\newP_i\ra N_j}

which, combined with $i,j$ above being algebra homomorphisms, gives
our $\kappa$-Poincare algebra as
\eqn{kpa}{ [P_0,\newP_i]=0,\quad [M_i,M_j]=\eps_{ijk}M_k,\quad
[\newN_i,\newN_j]=-\eps_{ijk}M_k}
\eqn{kpb}{[P_0,M_i]=0,\quad [\newP_i,M_j]=\eps_{ijk}\newP_k}
\eqn{kpc}{[P_0,\newN_i]=-\newP_i,\quad [\newP_i,\newN_j]=-\delta_{ij}({\kappa
\over
2}(1-e^{-{2P_0\over\kappa}})+{1\over 2\kappa}{\vec \newP}^2)+{1\over
\kappa}\newP_i\newP_j}
which is analagous to \cite{RT}. The coproducts become
\eqn{kpd}{ \Delta \newN_i=\newN_i\tens 1+e^{-{P_0\over\kappa}}\tens
\newN_i+{\eps_{ijk}\over \kappa}\newP_j\tens M_k,\quad \Delta M_i=M_i\tens
1+1\tens M_i.}
 In terms of $\newP_i$ the coproduct structure of $T$
itself is
\eqn{kpe}{\Delta P_0=P_0\tens 1+1\tens P_0,\quad \Delta \newP_i=\newP_i\tens
1+e^{-{P_0\over \kappa}}\tens \newP_i.}

Thus the new generators $\{P_0,\newP_i,\newN_i,M_i\}$
provide a natural description of $\kp $ as a Hopf algebra
bicrossproduct $U(so(1,3))\cobicross T$ according to the general construction
introduced in \cite{Ma:phy}.
The symbol $\cobicross$ denotes that one factor acts on the other and the
other coacts back on the first. Usually in the theory of groups and
Hopf algebras one considers only an action or coaction, but it was
argued in \cite{Ma:phy} that in physics actions tend to have
`reactions' and this turns out to be the case here when
$\kappa<\infty$.

Indeed, in \cite{Ma:pla}\cite{Ma:hop} one finds an example of the form
$U(su(2))\cobicross T$ where $T$ is the Hopf algebra of functions on $\R^3$
with a deformed coproduct corresponding to curvature from the point of view
there, and the action is a deformation of the usual rotations of $\R^3$. This
was one of the first non-trivial non-commutative non-cocommutative Hopf
algebras, though not as widely known as the celebrated Hopf algebras of
Drinfeld and Jimbo. The $\kp $ is quite similar to this but deformed in the
action of the boosts rather than of rotations.

\new{3.} We are now in a position to introduce a natural notion of
$\kappa$-Minkowski space on which our $\kp $ acts covariantly. Indeed,
since $T$ is the enveloping algebra of translations, it is natural to take for
$\kappa$-Minkowski its dual $T^*$ which will also be an algebra and on which
$T$ necessarily acts covariantly as quantum vector fields. We then show that
the whole of $\kp $ acts on it.

The structure of $T^*$ is completely determined by the axioms of a Hopf algebra
duality
\eqn{duality}{ <t,xy>=<t\o,x><t\t,y>,\quad <ts,x>=<t,x\o><s,x\t>,\quad \forall
t,s\in
T,\quad x,y\in T^*}
Indeed, since $T$ is the (commutative limit) of the borel subalgebra $U_q(b_-)$
of $U_q(su_2)$ and, as is well-known in that context, its dual is of the same
form\cite{Dri}. Thus, we take for $T^*$ the generators  $x_\mu$ and relations
and coproduct
\eqn{T*}{ [x_i,x_j]=0,\quad [x_i,x_0]={x_i\over\kappa},\quad \Delta
x_\mu=x_\mu\tens1+1\tens x_\mu.}
For $T$ we again prefer the generators $\newP_i, P_0$
and then the duality pairing can be written compactly as
\eqn{TT*}{ <f(\newP_i,P_0),:\psi(x_i,x_0):>=f({\del\over\del x_i},
{\del\over\del
x_0})\psi(0,0)}
where $:\psi(x_i,x_0):$ denotes a function $\psi$ of the generators {\em with
all powers of $x_0$ to the right}. One can see \cite{Ma:reg} for the usefulness
of this way of working with this kind of Hopf algebra. Apart from this
ordering,
we see that the pairing is completely along the classical lines of the pairing
of the enveloping algebra of $\R^4$ with the Hopf algebra of functions on
$\R^4$, which is by letting the translation generators act and evaluating at
zero.

Now the canonical action of $T$ on $T^*$ is
\eqn{regact}{t\la x=<x\o,t> x\t,\quad \forall x\in T^*,\quad t\in T}
which in our case works out as
\eqn{TT*act}{ \newP_i\la :\psi(x_i,x_0):=:{\del\over\del
x_i}\psi(x_i,x_0):,\quad  P_0\la
:\psi(x_i,x_0):=:{\del\over\del x_0}\psi(x_i,x_0):}
i.e. by the classical way but remembering the Wick-ordering.

Next, $U(so(1,3)$ also acts on $T^*$. This is because it acts from the right on
$T$ and this action therefore dualises to an action from the left on $T^*$:
\eqn{dualact}{ <t,h\la x>=<t\ra h,x>,\quad \forall t\in T,\quad h\in
U(so(1,3)),\quad x\in
T^*}
which computes in our case as
\eqn{LT*}{ M_i\la x_j=\eps_{ijk}x_k, \quad M_i\la x_0=0,\quad
N_i\la x_j=-\delta_{ij}x_0, \quad N_i\la x_0=-x_i. }

It is not obvious, but the general theory of bicrossproduct Hopf algebras
ensures that the canonical action of $T$ on itself by multiplication and
$U(so(1,3))$ by $\ra$ generates an action of the semidirect product algebra
$\kp $ on $T$. This therefore dualises to an action on $T^*$ generated by
the actions of these subalgebras. So $P_0,\newP_i$ as above and
$M_i,\newN_i$ acting like $M_i,N_i$ in (\ref{LT*}). are a
canonical representation of the
$\kp $ on $\kappa$-Minkowski. Their extension to products of the spacetime
co-ordinates is via the covariance condition (\ref{covact}) using the
coproducts $\Delta M_i$, $\Delta \newN_i$ etc. from (\ref{kpd})--(\ref{kpe}).
Thus,

\eqn{kpacta}{ M_i\la x_j=\eps_{ijk}x_k, \quad M_i\la x_0=0,\quad \newN_i\la
x_0=-x_i,\quad
\newN_i\la x_j=-\delta_{ij}x_0,}
\eqn{kpactb}{ \newN_i\la(x_jx_0)=-\delta_{ij}x^2_0-x_jx_i, }
\eqn{kpactc}{ \newN_i\la(x_0x_j)=-\delta_{ij}x^2_0-x_ix_j+{1\over
\kappa}\delta_{ij}x_0, }
\eqn{kpactd}{ \newN_i\la(x_jx_k)=-\delta_{ij}x_0x_k-\delta_{ik}x_jx_0+{1\over
\kappa}(\delta_{ik}x_j-\delta_{jk}x_i), }
\eqn{kpacte}{ \newN_i\la(x^2_0)=-x_ix_0-x_0x_i+{1\over \kappa}x_i. }
The Lorentz-invariant metric turns out as
\eqn{metric}{ x^2_0-{\vec x}^2+{3\over \kappa}x_0 }

This covariant action of  $\kp$ on $\kappa$-Minkowski space $T^*$ is our main
result of this section. It appears to be rather non-trivial to verify
it directly. Note that covariance is always true for $T$ on $T^*$ and since $T$
is a subhopf algebra of
$\kp $, it remains true as its translation sector. The classical boosts do
not act covariantly on $T^*$ but their coproduct is different in $\kp $ due
to the coaction $\beta$. This modification of the coproduct is just what is
needed for the
construction to work. The proof is straightforward using the abstract Hopf
algebra theory of Section~2.

We therefore have the correct basis for wave-functions $\psi$ on
$\kappa$-Minkowski space and can proceed with various constructions, retaining
at all time covariance under $\kp $. This will be explored elsewhere.

\new{4.} Our structure theorem for the $\kp $ has many other consequences
for the theory. The first of them is that the theory of bicrossproducts is
completely symmetric under the process of taking duals (reversing the roles of
products and coproducts). This remarkable
`input-output' symmetry was the main physical motivation for the introduction
of the bicrossproduct construction in \cite{Ma:phy}\cite{Ma:pla}\cite{Ma:phy}
and several other papers by the first author.

Thus we can compute the function algebra dual to $\kp $ at once. It is the
bicrossproduct
\eqn{funkp}{ \C(SO(1,3))\hookrightarrow T^*\bicross \C(SO(1,3)) \to T^*}
where $\C(SO(1,3))$ is the usual commutative algebra of functions on the
Lorentz group, and $T^*$ is our algebra of functions on
$\kappa$-Minkowski. Thus, this Hopf algebra is a deformation of the algebra of
functions on the Poincare group.
The maps and action/coaction for this dual construction are given in
\cite{Ma:phy} by dualising
the above $\beta,\ra$ respectively according to
\eqn{T*L*}{ <h,x\la \Lambda>=<\beta(h),x\tens \Lambda>,\quad \forall\ h\in
U(so(1,3)), \ x\in T^*,\ \Lambda\in \C(SO(1,3))}
\eqn{betaT*}{<t\tens h, \beta(x)>=<t\ra h,x>,\quad \forall t\in T,\ x\in T^*,\
h\in U(so(1,3)).}
The resulting $\kappa$-Poincare function Hopf algebra will be developed in
detail elsewhere. It can perhaps be compared with a $\kappa$-Poincare Hopf
algebra proposed in another context in \cite{Z}\cite{Zau}. In our approach it
necessarily
comes with a duality pairing with $\kp$ given by (\ref{TT*}), the usual pairing
between $\C(SO(1,3))$ and $U(so(1,3))$, and the trivial pairing (provided by
the counits) between translation and Lorentz sectors.

We conclude with some remarks about the physical interpretation of
bicrossproducts in \cite{Ma:pla} as quantum systems. Returning to our
enveloping algebra $\kp $ we can develop
quite a different physical picture. Namely, we think of $T$ not as the
enveloping
algebra of deformed translations but as the perfectly classical Hopf algebra of
{\em functions} on a classical nonAbelian group $X$,
\eqn{funx}{ T=\C(X) }
where $X$ is the group given by exponentiating the Lie algebra $\Xi$ defined by

\eqn{ksi}{  [x_i,x_0]={x_i\over\kappa}, \quad [x_i,x_j]=0.}

These are just the relations of $T^*$ in Section~3 but we think of them no
longer as generating the co-ordinates of some non-commutative space but as
generating a Lie algebra. It is easy to exponentiate the Lie algebra to
a group $X$ described as a subset of $\R^4$ with a $\kappa$-deformed
(non-Abelian) addition law. In other words, $\kappa$ controls now the curvature
of our space $X$. We take this $X$ as the position space (configuration space)
of a quantum system.

Next, the Lie algebra $\Xi$ and the Lie algebra $so(1,3)$ fit together to form
a `matched pair' of Lie algebras. The concept (due to the first author in
\cite{Ma:phy}\cite{Ma:pla}) is that each Lie algebra acts on the other in a
matching way. In our case  $so(1,3)$ acts by $\la$, say, on $\Xi$ via usual
infinitesimal Lorentz transformation and $\Xi$ acts back from the right by
dualising $\beta$ from (\ref{betaL}) according to the formula
\eqn{ksiact}{ \xi\ra x_\mu=<\beta(\xi),x_\mu\tens \id>,\quad \forall \xi\in
so(1,3),\
x_\mu\in \Xi }
remember that the output of $\beta$ has its first tensor factor in $T$, which
we evaluate against the generators $x_i,x_0$ using the pairing (\ref{TT*}). The
two actions fit together as required for a right-left matched pair:
\eqn{mpaira}{\xi\la[x_\mu,x_\nu] =[\xi\la x_\mu,x_\nu]+[x_\mu, \xi\la
x_\nu]+(\xi\ra x_\mu)\la x_\nu-(\xi\ra x_\nu)\la x_\mu}
\eqn{mpairb}{[\xi,\eta]\ra x_\mu=[\xi\ra x_\mu,\eta]+[\xi, \eta\ra
x_\mu]+\xi\ra(\eta\la x_\mu)-\eta\ra(\xi\la x_\mu)}
for all $\xi,\eta\in so(1,3)$. In our case, we can compute $\ra$ explicitly as
\eqn{xact}{ M_i\ra x_0=0 ,\quad M_i\ra x_j=0,\quad N_i\ra
x_0=-{1\over\kappa}N_i ,\quad N_i\ra x_j={1\over\kappa}\epsilon_{ijk}M_k }
and verify (\ref{mpaira})-(\ref{mpairb}) directly for these Lie algebra
representations $\la,\ra$. The $N_i,M_i$ here are the classical $so(1,3)$
generators .

The theory of such Lie algebras acting one eachother in such a way is a rich
one\cite{Ma:phy} and tells us among other things that there is a Lie algebra
double semidirect sum $\Xi\dcross so(1,3)$ containing $\Xi,so(1,3)$ and cross
relations
\eqn{xiL}{ [\xi,x_\mu]=\xi\la x_\mu + \xi\ra x_\mu.}
Moreover, there are theorems that, at least locally, the Lie algebra matched
pair exponentiates into a Lie group matched pair $X,SO(1,3)$ acting on each
other in a suitable way. The procedure and general formulae (which are
non-linear) have been introduced in \cite{Ma:mat}. There is also a double cross
product group $X\dcross SO(1,3)$, at least locally.

Now, the action of $SO(1,3)$ on $X$ has orbits. Consider particles constrained
to move on such orbits. The position obervables are $\C(X)$, the momentum
observables are the Lie algebra $so(1,3)$ since its elements generate the
flows. The natural quantisation of particles on such homogeneous spaces
according to the standard Mackey scheme\cite{Mac:ind}\cite{DobTol:mec} is
the cross product algebra $U(so(1,3))\rcross \C(X)$. This can be made precise
using the theory of $C^*$-algebras. The point is that this cross product
contains the algebra of $so(1,3)$ and $\C(X)$, with cross-relations which are
the natural covariant form of Heisenberg's commutation relations. Our
$\kp $ is this quantum algebra of observables.

Moreover, the dual of the bicrossproduct is also a bicrossproduct: it is the
quantisation of particles moving on the homogeneous spaces which are the orbits
in $SO(1,3)$ under the action of $X$, i.e. precisely with the roles of position
and momentum reversed. Thus models of this class, demonstrated here by
$\kp $ exhibit a quantum version of Born reciprocity and are interesting
for this reason\cite{Ma:pla}\cite{Ma:phy}. Moreover, this structure  generally
forces the action to be deformed, often with event-horizon-like singularities.
For example, it was shown in \cite{Ma:pla} that the extensions of
$\C(\R\times\R)$ (the classical phase-space in one-dimensions) of this
bicrossproduct type had just two free parameters, which we identified
heuristically as $\hbar$ and $G$, the gravitational coupling constant. This
work was perhaps one of the first serious attempts to apply Hopf algebras and
non-commutative geometry to Planck scale physics, and it is interesting that
$\kp $ has an interpretation in these terms as well as a symmetry in
particle physics. This
picture of the $\kappa$-Poincare algebra will be developed in detail elsewhere.


\end{document}